# Two-dimensional Platinum Diselenide Waveguide-Integrated Infrared Photodetectors

*Shayan Parhizkar[1,2], Maximilian Prechtl[3], Anna-Lena Giesecke[2], Stephan Suckow[2,*], Sophia Wahl[4], Sebastian Lukas[1], Oliver Hartwig[3], Nour Negm[1,2], Arne Quellmalz[5], Kristinn Gylfason[5], Daniel Schall[2,6], Matthias Wuttig[4], Georg S. Duesberg[3], and Max C. Lemme[1,2,*]*

[1] Chair of Electronic Devices, RWTH Aachen University, Otto-Blumenthal-Str. 2, 52074 Aachen, Germany

[2] AMO GmbH, Advanced Microelectronic Center Aachen, Otto-Blumenthal-Str. 25, 52074 Aachen, Germany

[3] Institute of Physics, Faculty of Electrical Engineering and Information Technology (EIT 2) and Center for Integrated Sensor Systems, University of the Bundeswehr Munich, 85577 Neubiberg, Germany

[4] Institute of Physics IA, RWTH Aachen University, 52074 Aachen, Germany

[5] Division of Micro and Nanosystems, School of Electrical Engineering and Computer Science, KTH Royal Institute of Technology, SE-10044 Stockholm, Sweden

[6] Black Semiconductor GmbH, Schloss-Rahe-Str. 15, 52072 Aachen Germany

Abstract

Low cost, easily integrable photodetectors (PDs) for silicon (Si) photonics are still a bottleneck for photonic integrated circuits (PICs), especially for wavelengths above 1.8 $\mu$m. Multilayered platinum diselenide ($PtSe_2$) is a semi-metallic two-dimensional (2D) material that can be synthesized below 450°C. We integrate $PtSe_2$ based PDs directly by conformal growth on Si waveguides. The PDs operate at 1550 nm wavelength with a maximum responsivity of 11 mA/W and response times below 8.4 $\mu$s. Fourier transform infrared spectroscopy (FTIR) in the wavelength range from 1.25 $\mu$m to 28 $\mu$m indicates the suitability of $PtSe_2$ for PDs far into the infrared wavelength range. Our $PtSe_2$ PDs integrated by direct growth outperform $PtSe_2$ PDs manufactured by standard 2D layer transfer. The combination of IR responsivity, chemical stability, selective and conformal growth at low temperatures, and the potential for high carrier mobility, make $PtSe_2$ an attractive 2D material for optoelectronics and PICs.

KEYWORDS: platinum diselenide, photodetector, silicon photonics, 2-dimensional materials, infrared

Introduction

Photonic integrated circuits (PICs) are maturing as a platform for applications in telecommunications, spectroscopy, diagnostics, biomedical imaging, and gas sensing.[1] PICs tailored for these applications typically rely on infrared (IR) photodetectors that are coupled to waveguides and that function as essential components for optoelectronic signal conversion. Photodetectors for the near-infrared wavelength range are mainly based on widely used semiconductors such as Si[2] and Ge[3]. Mid-infrared photodetectors are typically made from compound semiconductors such as InGaAs[4] and HgCdTe[5], with the downside of higher manufacturing cost and the necessity of cryogenic temperature operating conditions. However, a key requirement towards broad applicability of PICs is the integrability of IR photodetectors on photonic waveguides. This is not fulfilled by commercial manufacturing methods for the conventional materials, because their high deposition and (epitaxial) growth temperatures are not compatible with PIC thermal budgets. This creates a demand for infrared photodetectors with high manufacturability at low temperature budgets.

Two-dimensional (2D) materials have attracted tremendous attention in the optoelectronics field due to their broadband optical absorption, high carrier mobility, mechanical flexibility and ease of integration.[6–10] Quantum confinement in 2D materials in the direction perpendicular to their 2D plane leads to novel physical properties that distinguish them from their bulk materials.[11] A wide range of devices like broadband photodetectors,[12] modulators,[13,14] Lasers,[15] light emitting diodes,[16] phototransistors,[17–19] and avalanche photodiodes[20] have been demonstrated that emphasize the advantages of 2D materials for optoelectronics.

In particular, the zero-bandgap material graphene has been used to demonstrate high performance integrated photodetectors and modulators in the IR.[21–25] However, graphene requires a layer transfer process,[26,27] because its growth is limited to a few substrate materials and requires high temperatures. 2D black phosphorus is also suitable as an IR PD[28,29] material that can be transferred onto waveguides,[30,31] but has the drawback that it is not entirely stable in ambient conditions.[32] Platinum diselenide ($PtSe_2$) is a transition metal dichalcogenide (TMD) with an octahedral lattice structure that is semiconducting with an indirect bandgap of 1.2 eV as a monolayer. As a multilayer material, it becomes semi-metallic.[33] This semi-metallic nature of layered $PtSe_2$ allows its use for broadband IR-photodetection in a similar way to graphene.[34–36] In addition, $PtSe_2$ is stable in air and has a high charge carrier mobility compared to other TMDs, with theoretically predicted values of more than 1,000 $cm^2/Vs$,[37,38] and has a piezoresistive gauge factor of up to -85.[39–41] A major advantage of $PtSe_2$ is the possibility of direct, large-scale growth on various substrates at temperatures below 450°C using the thermally assisted conversion (TAC) technique.[34,40,42,43] In combination with the potential for selective and conformal deposition,[44] $PtSe_2$ is suitable for back-end-of-line (BEOL) integration on electronic and photonic wafers.

In this work, we demonstrate the use of layered $PtSe_2$ as integrated IR photodetectors on Si photonic waveguides. We compare the device performance and material quality of a direct-growth integration approach with a wet layer transfer technique through analytical, electrical, and optical characterization.

Experimental:

The $PtSe_2$ photodetectors were realized on rib-waveguides with 50 nm step height on silicon on insulator (SOI) substrates. Optical access is provided through two grating couplers that are optimized for a wavelength of 1550 nm and transverse electric (TE) polarization. We fabricated two sets of samples, one using a direct-growth and a second one using a wet-transfer method of the PtSe$_2$ layers. For the first set, a 10.8 nm thick layer of pre-patterned sputtered platinum (Pt) was converted into a 27 nm thick layered film of PtSe$_2$ directly on the waveguides using TAC.[34] Pt is a highly reactive material. A thin barrier layer of Al$_2$O$_3$ was therefore deposited with atomic layer deposition (ALD) to protect the waveguides and grating couplers from the reaction of Pt with the top Si photonic layer during the TAC growth. For the second set of samples, PtSe$_2$ films of 7.6, 13.7, and 23.5 nm thickness were grown by TAC on separate silicon substrates with a 90 nm thermal silicon oxide (SiO$_2$) layer and were wet transferred onto the Si waveguides (see Methods for details). The thicknesses of the PtSe$_2$ films were measured by atomic force microscopy (AFM) (supporting information). In both sample sets, the PtSe$_2$ patches were contacted with nickel/aluminum (Ni/Al) electrodes that had a distance of 5 μm to the waveguides. The width of the PtSe$_2$ channels on the waveguide along the light propagation direction for all photodetectors is W = 50 μm. Schematic cross sections of the integrated photodetectors and scanning electron microscope (SEM) images of a transferred and a directly grown photodetector are shown in Fig. 1 a-d.

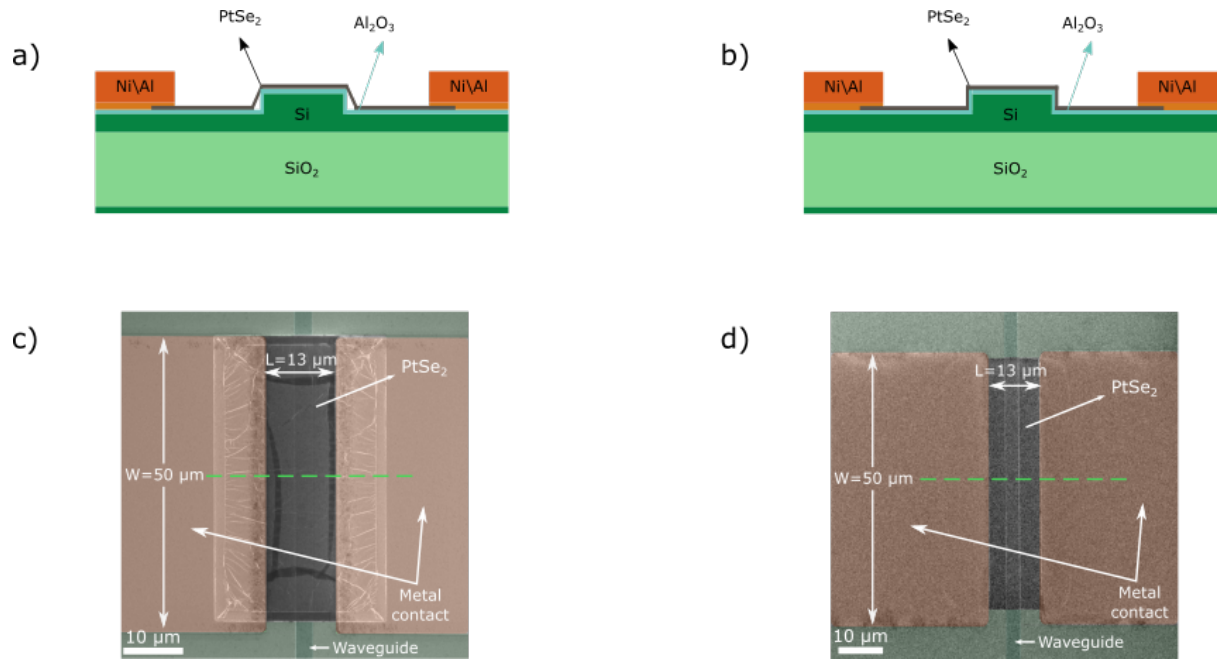

Figure 1. Schematic cross sections and false color scanning electron microscope (SEM) micrographs of transferred (a, c) and directly grown (b,d) PtSe$_2$ photodetectors on silicon waveguides. The wrinkles visible in the SEM image in c) are due to wet transfer, and consequently absent in d). The green dashed lines indicate the direction of the cross sections in a) and b).

The topography of the PtSe$_2$ films around the waveguides is different for the two different integration methods. The wet transfer method presumably leads to small air gaps around the sidewalls of the Si waveguides, which is known from graphene devices,[45] as indicated schematically in Fig. 1a. With the wet transfer process, the conformal coverage around the waveguide is not possible because of the limited flexibility of the transfer polymer. In addition, wrinkles in the PtSe$_2$ layer are visible that result from the wet transfer process (Fig. 1c). In the case of directly grown PtSe$_2$, the sputtered Pt covers the sidewalls of the waveguides which results in a conformal PtSe$_2$ film around the waveguide (Fig. 1b and d). This feature underlines the advantage of direct 2D material growth over layer transfer which often requires additional planarization processes to avoid tearing of the 2D films at the sharp edges of the waveguides[24,46] and reduces the strain effects arising from the bending of the 2D films over the waveguides.[47,48]

Raman spectroscopy was carried out on all devices to confirm the successful formation and quality of PtSe$_2$ layers after fabrication (Fig. 2a). The spectra of various PtSe$_2$ films on silicon-on-insulator substrate exhibit two characteristic peaks at approximately 177 cm$^{-1}$ and 206 cm$^{-1}$, which represent the E$_g$ and A$_{1g}$ modes of layered PtSe$_2$, respectively.[49] The E$_g$ peak originates from the in-plane vibration of selenium (Se) atoms and the A$_{1g}$ peak is caused by out of plane vibration of Se atoms. As the number of PtSe$_2$ layers increases, a red shift in the position of both peaks and an increase in the intensity ratio of the two peaks $I(A_{1g})/I(E_g)$ is observed. This behavior can be explained by an increasing out-of-plane contribution due to an increase of Van-der-Waals interactions between the layers.[49] The full width at half maximum (FWHM) of the E$_g$ peak indicates the material quality of PtSe$_2$.[50,51] For high quality TAC-grown PtSe$_2$ films, the FWHM is smaller than 5 cm$^{-1}$.[41] For our samples with a directly grown PtSe$_2$ film the FWHM of the E$_g$ peak is 4.8 cm$^{-1}$. For the samples with transferred PtSe$_2$, this value varies between 5.1 and 5.6 cm$^{-1}$. In both cases, these values indicate high material quality which is sufficient for device integration.[41] We note that there is still a quality gap between highly crystalline exfoliated films and thin films grown by various techniques.

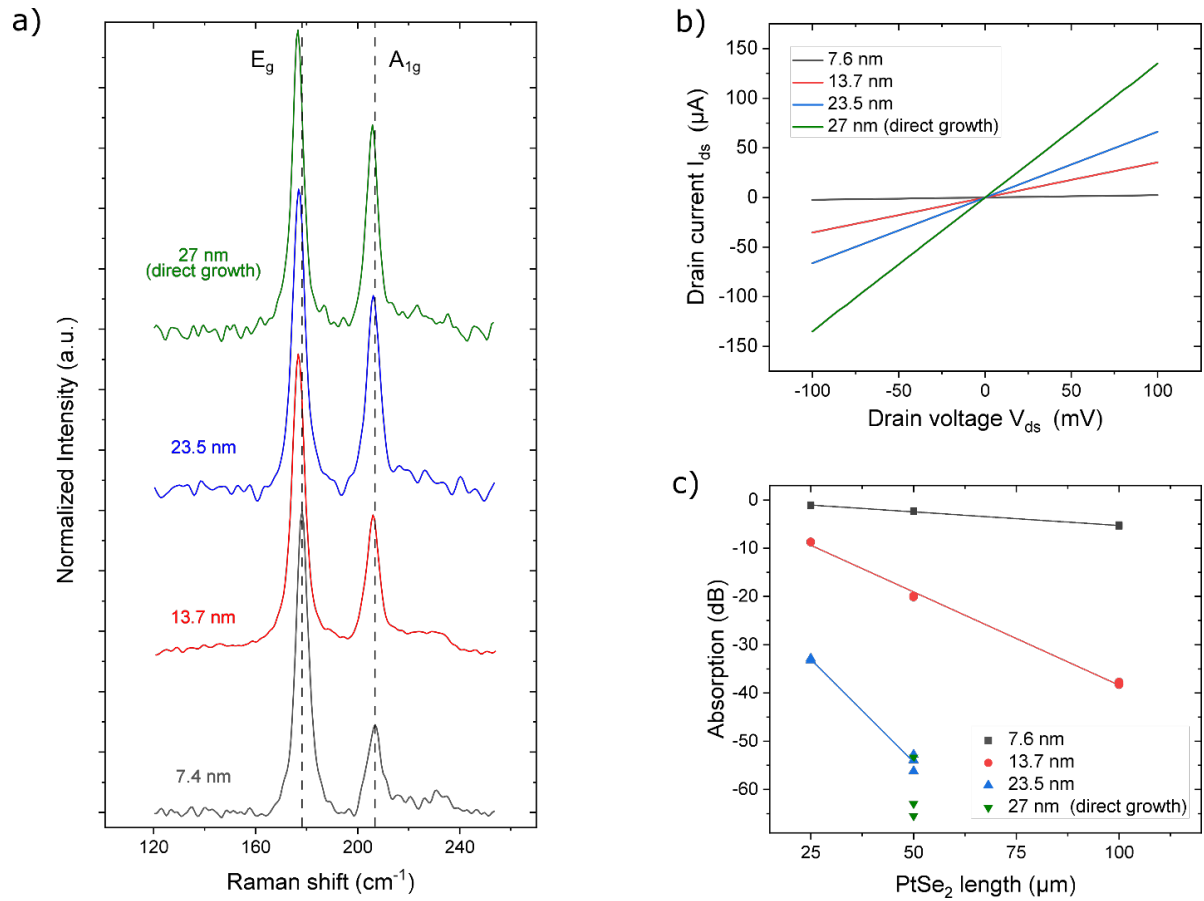

Figure 2. a) Raman spectra of PtSe$_2$ with different thicknesses of transferred PtSe$_2$ (gray, red, and blue lines) and direct growth (green line). The spectra represent the average of area scans. They were recorded with an integration time of 1 s and averaged over 10 accumulations. b) Drain current ($I_{ds}$) as a function of drain-source voltage ($V_{ds}$) for different PtSe$_2$ photodetectors. c) Evanescent field absorption of PtSe$_2$ films on the waveguide at 1550 nm wavelength. The absorption per propagation distance for 7.6 nm, 13.7 nm, 23.5 nm, and 27 nm thick (directly grown) PtSe$_2$ is 0.06, 0.38, 0.85, and 1.2 dB/μm.

Current-voltage (IV) measurements of the PtSe$_2$ detectors with different thicknesses show near-linear behavior for small source-drain voltages, which indicates semi-metallic characteristics of the PtSe$_2$ films and ohmic contacts to the Ni/Al electrodes (Fig 2b). The resistance of $PtSe_2$ decreases with increasing layer thickness. The directly grown $PtSe_2$ layer exhibits similar properties as the transferred layers. However, as the drain voltages are increased, the IV curves clearly deviate from the linear behavior, which indicates the existence of back-to-back Schottky barriers at the interfaces (Fig S3 in the supporting information). This is confirmed by taking the derivative of the IV curves, which show that a weak nonlinearity is also present for smaller bias voltages and slightly stronger for thinner films (Fig S4 in the supporting information). The latter is in line with previous reports that have observed semiconducting behavior for thin TAC films.[43] These films are still multilayer films, and the reason of their semiconducting behavior is under discussion, because in theory, the semi-metal to semiconductor transition should happen predominantly for monolayer films.[52] The signatures of Schottky barriers at the contact interfaces of thicker films, which are expected to be semi-metallic, has been explained by the coexistence of semi-metallic and semiconducting domains in the TAC grown films.[53] Nevertheless, the resistivities of nickel/aluminum contacts to $PtSe_2$ films have been measured with the transfer length method (TLM) to vary between 0.7 and 2 k$\Omega\,\mu$m for the same batch of samples in a previous study.[41] Such values allow us to neglect the effect of the metal contacts on the total device resistance for our purposes and to use the two-probe configuration for further analysis. This is an important observation because the direct growth process includes Pt deposition and TAC on the sidewalls and at the edges and corners, which is the key enabler for the direct integration of $PtSe_2$ on photonic structures.

The light-matter interaction of the $PtSe_2$ films was characterized for infrared light with a wavelength of $\lambda = 1550$ nm that was coupled into one grating coupler through a single-mode fiber. The fiber-to-fiber losses of the grating couplers and the waveguides of 8 dB were measured on separate test structures and subtracted to obtain these values. First, we measured the specific absorption of the evanescent field by the PtSe$_2$ films located at 10 nm distance to the waveguide, using structures with different PtSe$_2$ dimensions (Fig. 2c). The resulting absorption was 0.06, 0.38, 0.85, and 1.2 dB/$\mu$m for 7.6, 13.7, 23.5 and 27 nm PtSe$_2$, respectively.

The opto-electric response of the PtSe$_2$ photodetectors was investigated by measuring the photocurrents with a lock-in amplifier, while modulating the light intensity at a frequency of 1 kHz (see also supporting information). The photocurrents were measured as a function of laser light power for each photodetector with a width of 50 $\mu$m (along the waveguides) and a length of 13 $\mu$m (perpendicular to the waveguides, i.e. distance between the contacts) for an applied bias voltage of $V_{ds}$ = 4.5 V across the device. All detectors exhibit a linear dependence on light power for the measured range. The highest photocurrent is observed for the device with the thickest PtSe$_2$ film which is directly grown on the waveguide (Fig. 3a). The intrinsic responsivity is defined as $R = I_{ph}/P_{opt}$, where $I_{ph}$ is the photocurrent and $P_{opt}$ is the optical power arriving at the photodetector after subtracting the losses of the grating couplers and waveguides. Measurements of the responsivities at different bias voltages ranging from $V_{ds}$ = -4.5 V to 4.5 V at a fixed laser output power of 9 mW are plotted in Fig. 3b. Increasing $V_{ds}$ results in higher photocurrents and consequently higher responsivities. At 4.5 V applied bias voltage, this intrinsic responsivity yields $R$ = 0.8 mA/W for the thinnest PtSe$_2$ layer (7.6 nm) up to a maximum value of $R$ = 11 mA/W for the photodetector with a directly grown PtSe$_2$ layer of 27 nm thickness (Table 1). Different devices on each chip were measured and statistics of the responsivities are also shown in Fig. 3c, which confirms that the responsivity of the devices increases with the number of PtSe$_2$ layers.

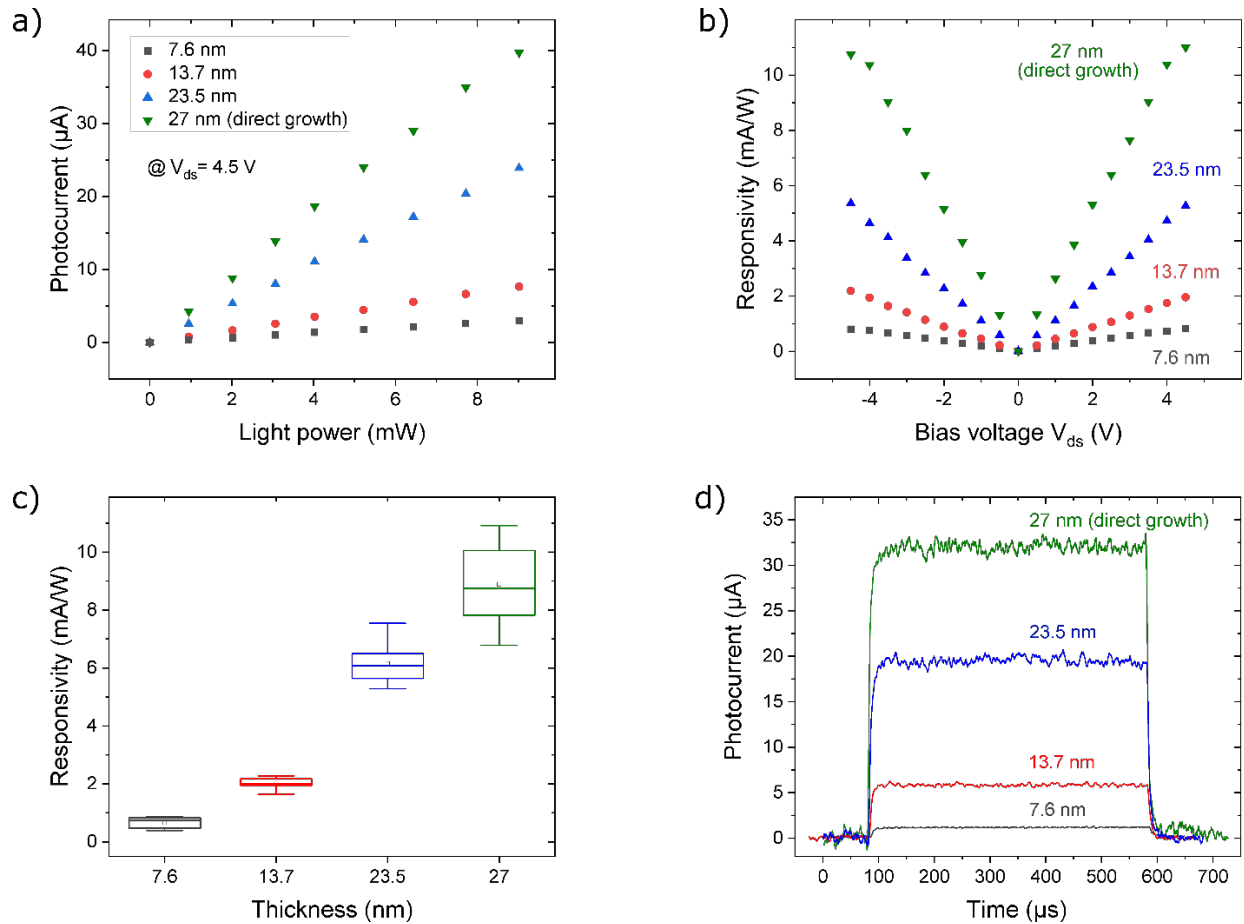

Figure 3. Optical measurements of PtSe$_2$ photodetectors. All devices are 50 $\mu$m wide along the waveguide direction. a) Photocurrent as a function of light power at 4.5 V bias voltage. All PtSe$_2$ photodetectors respond linearly to the light power. b) Responsivity versus bias voltage calculated at 9 mW laser output power. c) Box plot of responsivities for different thickness PtSe$_2$ photodetectors measured at 4.5 V applied bias. d) Time resolved measurement of the PtSe$_2$ photodetectors at 8 mW optical power and 4.5 V bias voltage. The measured rise and fall times for the devices are between 8 and 13 $\mu$s.

The normalized photocurrent to dark current ratio (NPDR), defined as responsivity divided by the dark current, is another important parameter for evaluating the sensitivity of the photodetectors. Larger NPDR values indicate better suppression of dark currents and lower noise-equivalent power. NPDRs of all the photodetectors have been calculated and listed in Table 1. The NPDR values for the PtSe$_2$ photodetectors are similar to those reported for graphene photodetectors but have the strong advantage of enabling direct growth instead of transfer.[22,54]

Time-resolved measurements were performed on the PtSe$_2$ photodetectors (Fig. 3d). The rise and fall time are defined by the time it takes for the photocurrent to reach 10 % and 90 % of the maximum value in rising and decaying curves, respectively, or by fitting equations 1 and 2:[55]

$$I_{rise} = I_0 + Ae^{\left(\frac{t-t_1}{\tau}\right)} \qquad (eq.\ 1)$$

$$I_{fall} = I_0 + Be^{\left(-\frac{t-t_2}{\tau}\right)} \qquad (eq.\ 2)$$

where $\tau$ is the rise/fall time constant and $t_1$ or $t_2$ is the time, it takes for switching the laser on or off, respectively. The measured rise and fall times for the photodetectors are listed in table 1. The values for all the detectors are between 8 and 13 $\mu$s. Consequently, our photodetectors provide a faster response time than most of the reported TMD-based IR photodetectors so far.[56,57] The performance parameters of merit for some 2D-based infrared photodetectors are summarized in Table 2. The transient time ($\tau_t$), i.e. the time needed for the photocarriers to reach the metal contacts, can be calculated through the carrier drift model by equation 3:[58]

$$\tau_t = \frac{L^2}{2\mu V_{ds}} \qquad (eq.3)$$

where $L$ is the length of the channel perpendicular to the propagation of the light, $\mu$ is the electron mobility and $V_{ds}$ is the applied bias. To approximate the carrier mobility, PtSe$_2$ films with the same thicknesses as used for the photodetectors and grown in the same batch were wet transferred onto a Si substrate with 90 nm SiO$_2$ layer. After fabrication of the contacts and patterning of the PSe$_2$ films, the field effect mobility of all the film thicknesses were calculated using the transconductance method in equation 4:[59,60]

$$\mu = g_m \frac{L}{W V_{ds} C_g} \qquad (eq.\ 4)$$

Here, W is the channel width along the waveguide and $V_{ds}$ is the voltage applied to the device channel. $g_m$ is defined as $\frac{dI_{ds}}{dV_g}$ where $I_{ds}$ is the drain-source current and $V_g$ is the back gate voltage, and $C_g = \frac{\varepsilon \varepsilon_0}{t_{ox}}$ is the back gate capacitance of SiO$_2$. $\varepsilon$ and $\varepsilon_0$ are the relative and vacuum permittivity and $t_{ox}$ is the thickness of SiO$_2$. The extracted field effect mobilities for all PtSe$_2$ films were between 2.7 to 3.9 cm$^2$/Vs. This relatively low value can be attributed to defects and the polycrystallinity of the films grown by the TAC method.[39,47] Literature data on mobilities of PtSe$_2$ films show broad variability ranging from values below 1 cm$^2$/Vs for TAC grown films[42,61,62] to 210 cm$^2$/Vs for exfoliated films.[38] Most of the studies report mobilities below 50 cm$^2$/Vs.[63–67] Using the extracted carrier mobility, the transient time for a device with $L = 13\ \mu$m and 4.5 V bias voltage is 0.6 ns.

Table 1. Summary of the different PtSe$_2$ photodetectors. Maximum responsivity, normalized photocurrent to dark current ratio (NPDR) and rise/fall time of all the photodetectors.

| PtSe$_2$ | Thickness (nm)[a] | Number of layers (estimate) | Absorption (dB/μm) | Max responsivity (mA/W)[b] | NPDR (W$^{-1}$)[b] | Rise/fall time (μs)[b] |
|---|---|---|---|---|---|---|
| Sample A: Grown on Si/SiO$_2$, wet transferred | 7.6 | 11 | 0.6 | 0.87 | 7.5 | 8.6/13.1 |
| Sample B: Grown on Si/SiO$_2$, wet transferred | 13.7 | 20 | 0.38 | 2.2 | 1.3 | 8.7/9.9 |
| Sample C: Grown on Si/SiO$_2$, wet transferred | 23.5 | 34 | 0.85 | 7.5 | 1.6 | 8.5/9.7 |
| Sample D: Directly grown on Si waveguide | 27 | 39 | 1.2 | 11 | 1.9 | 8.4/8.7 |

(a) A single layer PtSe$_2$ has a thickness of about 0.7 nm.[68] Number of layers for each film can be estimated accordingly.

(b) Reported at 4.5 V bias voltage

Potential applications for photonics-integrated IR photodetectors beyond the telecommunication wavelengths are sensing, diagnostics, thermal imaging, and free space communication.[60,69] Therefore, we have studied the broad-band absorption of PtSe$_2$ films with different thicknesses for the wavelength range from 1.2 to 28 μm using Fourier transform infrared spectroscopy (FTIR). We wet transferred PtSe$_2$ films with thicknesses of 7.6, 13.7, and 23.5 nm onto separate Si substrates and measured their absorbance (A) by FTIR spectroscopy, while subtracting the absorbance of the Si substrate (Fig. 4a). All PtSe$_2$ films exhibit a small peak at 1729 cm$^{-1}$ (5784 nm). The absorption coefficients ($\alpha$) of the PtSe$_2$ films were also calculated from their absorbance by equation 5,

$$A \ln(10) = \alpha t \qquad (eq.\ 5)$$

where *A* is the absorbance of the material and *t* is the film's thickness (Fig. 4b). The absorption coefficient of the sample with the thickest PtSe$_2$ (23.5 nm) layer behaves differently from the other two samples: while its absorption decreases at first in a similar way as the 7.6 and 13.7 nm thick films, it starts to increase to a steady value as the wavelength increases further. The coefficients of the thinner films remain low as the wavelength increases. The absorption of the 23.5 nm thick PtSe$_2$ is non-zero even at 28 $\mu$m wavelength (0.04 eV) which can be considered metallic behavior.[70] This result suggests that PtSe$_2$ may also be suitable for long wavelength mid-IR photodetectors.

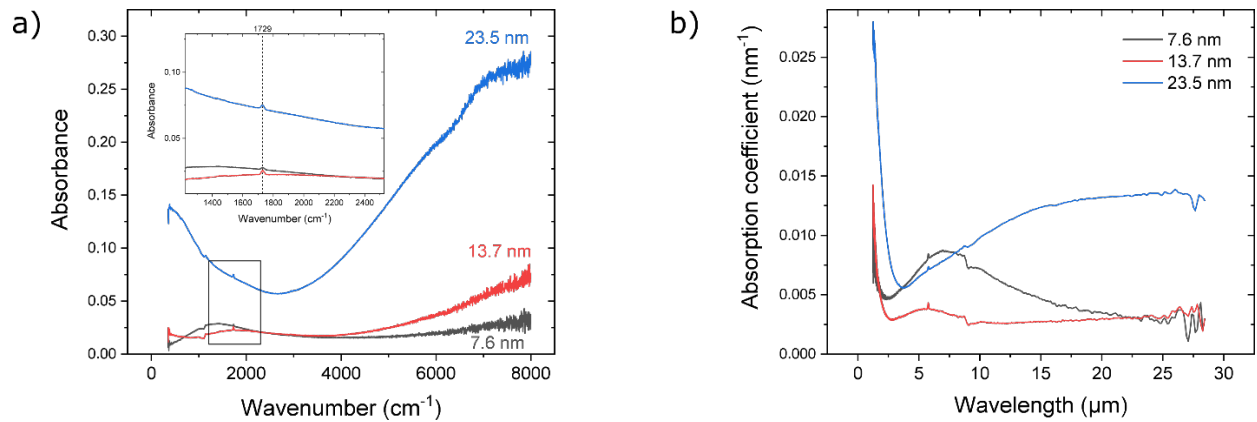

Figure 4. a) Absorbance of PtSe$_2$ by the use FTIR spectroscopy. Inset: Magnified view of the peak at 1729 cm$^{-1}$ (5784 nm). b) Absorption coefficients of PtSe$_2$ films with different thicknesses, calculated from their absorbance.

We have demonstrated the integration of PtSe$_2$ photodetectors into silicon photonics waveguides and demonstrated their functionality at a wavelength of 1550 nm. We have shown that layered PtSe$_2$ can be utilized for infrared photodetection with high responsivity. Our PtSe$_2$ photodetectors can be synthesized directly on the photonic waveguide structures, on wafer scale and at CMOS compatible temperatures using TAC. The directly grown films show higher performance than reference devices transferred with a typical 2D layer transfer method. The highest responsivity of 11 mA/W was achieved for a directly grown PtSe$_2$ photodetector, which reached a fast response time of 8.4 $\mu$s. FTIR data indicate that PtSe$_2$ is suitable for photodetection well into the mid-IR range, which opens opportunities for applications in food safety, agriculture, gas detection, on-chip spectroscopy, or imaging. The direct growth of PtSe$_2$ on waveguides creates a new perspective for the integration of 2D materials with photonic integrated circuits. Our results show that multi-layered PtSe$_2$ is a promising candidate for high-responsivity optoelectronic applications in the near and mid IR regime, including the direct integration on commercial semiconductor technology platforms. This includes silicon nitride (SiN) photonics, as the growth methods does not require any crystalline surfaces, in contrast to epitaxial processes.[40,66,71,72]

Table 2. Characteristics and performance metrics of photodetectors based on 2D materials.

| Materials | Growth methods | Wavelength | Responsivity | Response time | Ref. |
|---|---|---|---|---|---|
| Graphene | CVD | 1550 nm | 180 mA/W | 2.7 ps* | [73] |
| BP | Exfoliation | 3.8 $\mu$m | 11.3 A/W | 0.3 ms | [74] |
| MoS$_2$ | MOVPE | 360 - 960 nm | 920 A/W | 0.5/1.15 s | [57] |
| MoS$_2$ | Exfoliation | 980 nm | 2.3 A/W | 50 ms | [19] |
| HfS$_2$ | CVD | 808 | $3.8 \times 10^5$ A/W | 8 ms | [75] |
| MoTe$_2$ | CVD | 980 | 6.4 A/W | 31/21 ms | [76] |
| MoTe$_2$/Gr | Exfoliation | 1300 nm | 0.2 A/W | 7 ps* | [77] |
| PtSe$_2$ | TAC | 635 nm – 2.7 $\mu$m | 7.8 mA/W | 6/9 $\mu$s | [36] |
| PtSe$_2$ | CVD | 1550 nm | 0.19 mA/W | 17/39 ps | [78] |
| PtSe$_2$ | TAC | 1550 nm | 11 mA/W | 8.4/8.7 $\mu$s | This work |

* Calculated from reported 3 dB bandwidth.

Methods

Device Fabrication: silicon rib waveguides with 50 nm step height were fabricated on a 150 mm SOI wafer with 220 nm top Si and 3 $\mu$m buried oxide layers using i-line (365 nm) photolithography and reactive ion etching (RIE). Grating couplers designed for 1550 nm wavelength were realized at the end of the waveguides by electron beam lithography and subsequent RIE. 10 nm of $Al_2O_3$ were deposited on the wafer by ALD to protect the waveguides and grating couplers from diffusion of Pt into the top Si layer during the growth process and also to avoid the formation of a Schottky junction between $PtSe_2$ and the bottom Si. The thin $Al_2O_3$ works as a cladding for grating couplers and increases their coupling efficiency.

The SOI wafer was diced after fabricating the photonic base components and $PtSe_2$ photodetectors were fabricated on different dies of the wafer.

For the directly grown sample, a 10.8 nm Pt layer was sputtered onto the waveguides using a pre-defined lithography pattern and a lift-off process. The Pt layer was converted into $PtSe_2$ of 27 nm thickness. $PtSe_2$ films with thicknesses of 7.6, 13.7, and 23.5 nm were grown on separate $Si/SiO_2$ (90 nm) substrates and then wet transferred onto the separate samples using potassium hydroxide (KOH) solution. All $PtSe_2$ films were grown using this TAC process, described in detail in previous publications.[34,49] For wet transfer a support layer of Poly (methyl methacrylate) (PMMA) was spincoated on top of the $PtSe_2$ films. Then PMMA film was scratched, and few droplets of KOH solution were dropped on the scratched areas. KOH causes delamination of $PtSe_2$ films from underlying $SiO_2$. After delamination and release of $PtSe_2$/PMMA films from the substrates, they remained floating on the deionized water for few days and then were transferred on the final substrates using fishing technique. The samples were dried in air and then PMMA layer was removed from their surfaces using acetone and isopropanol. Transferred $PtSe_2$ films were patterned using contact lithography and reactive ion etching. Afterwards, all PtSe$_2$ films were contacted with 15 nm Ni and 50 nm Al using contact lithography and subsequently lift off process.

Electrical characterization: Electrical measurements were performed in a Lakeshore chamber connected to a Keithley SCS4200 source meter unit at ambient conditions.

FTIR measurements: The absorption spectra were calculated from transmittance measurements. The data was recorded from 0.05 eV (400 cm$^{-1}$) up to 0.99 eV (8000 cm$^{-1}$) in a Bruker Vertex 80v Fourier-transform spectrometer with a spectral resolution of 0.5 meV (4 cm$^{-1}$) and averaged over 64 scans.

ASSOCIATED CONTENT

Supporting Information

Thickness measurements of PtSe$_2$ films using atomic force microscopy, more details about opto-electrical measurements and set up, comments on IV curve characteristics

AUTHOR INFORMATION

**Corresponding Author**

* suckow@amo.de, max.lemme@eld.rwth-aachen.de

FUNDING SOURCES

This work has received funding from the European Union's Horizon 2020 research and innovation programme under grant agreements 825272 (ULISSES) and 881603 (Graphene Flagship), as well as the German Ministry of Education and Research (BMBF) under grant agreement 16ES1121 (NobleNEMS).

**TOC Graphic**

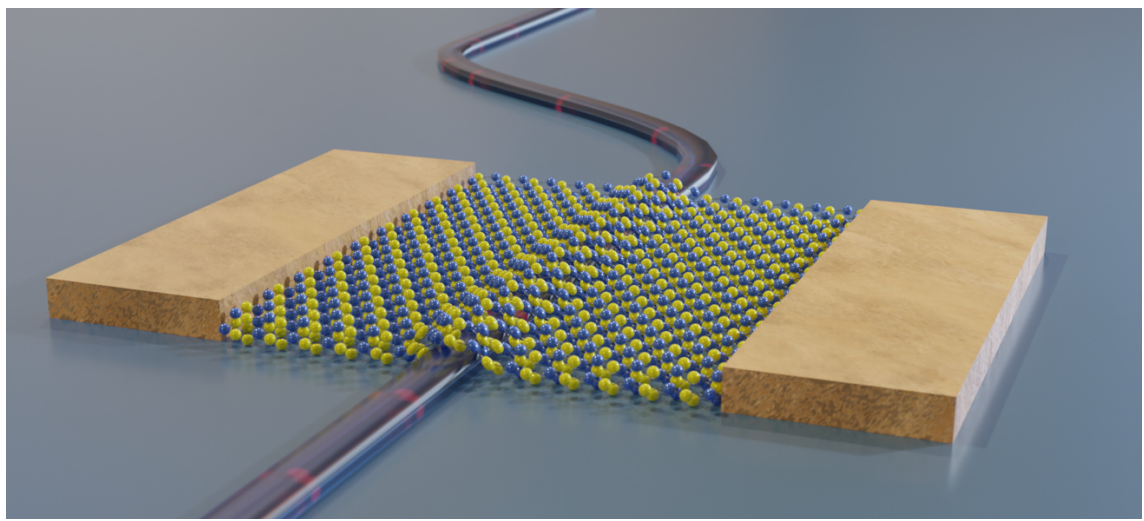

# Supporting Information for "Two-dimensional Platinum Diselenide Waveguide-Integrated Infrared Photodetectors"

*Shayan Parhizkar[1,2], Maximilian Prechtl[3], Anna Lena Giesecke[2], Stephan Suckow[2,\*], Sophia Wahl[4], Sebastian Lukas[1], Oliver Hartwig[3], Nour Negm[1,2], Arne Quellmalz[5], Kristinn Gylfason[5], Daniel Schall[2,6], Matthias Wuttig[4], Georg S. Duesberg[3], Max C. Lemme[1,2,\*]*

[1] Chair of Electronic Devices, RWTH Aachen University, Otto-Blumenthal-Str. 2, 52074 Aachen, Germany

[2] AMO GmbH, Advanced Microelectronic Center Aachen, Otto-Blumenthal-Str. 25, 52074 Aachen, Germany

[3] Institute of Physics, Faculty of Electrical Engineering and Information Technology (EIT 2) and Center for Integrated Sensor Systems, University of the Bundeswehr Munich, 85577 Neubiberg, Germany

[4] Institute of Physics IA, RWTH Aachen University, 52074 Aachen, Germany

[5] Division of Micro and Nanosystems, School of Electrical Engineering and Computer Science, KTH Royal Institute of Technology, SE-10044 Stockholm, Sweden

[6] Black Semiconductor GmbH, Schloss-Rahe-Str. 15, 52072 Aachen Germany

\*Email: suckow@amo.de, max.lemme@eld.rwth-aachen.de

**Thickness measurements of PtSe$_2$ films**

Thicknesses of all PtSe$_2$ films were measured by atomic force microscopy (AFM) and they are shown in Fig. S1.

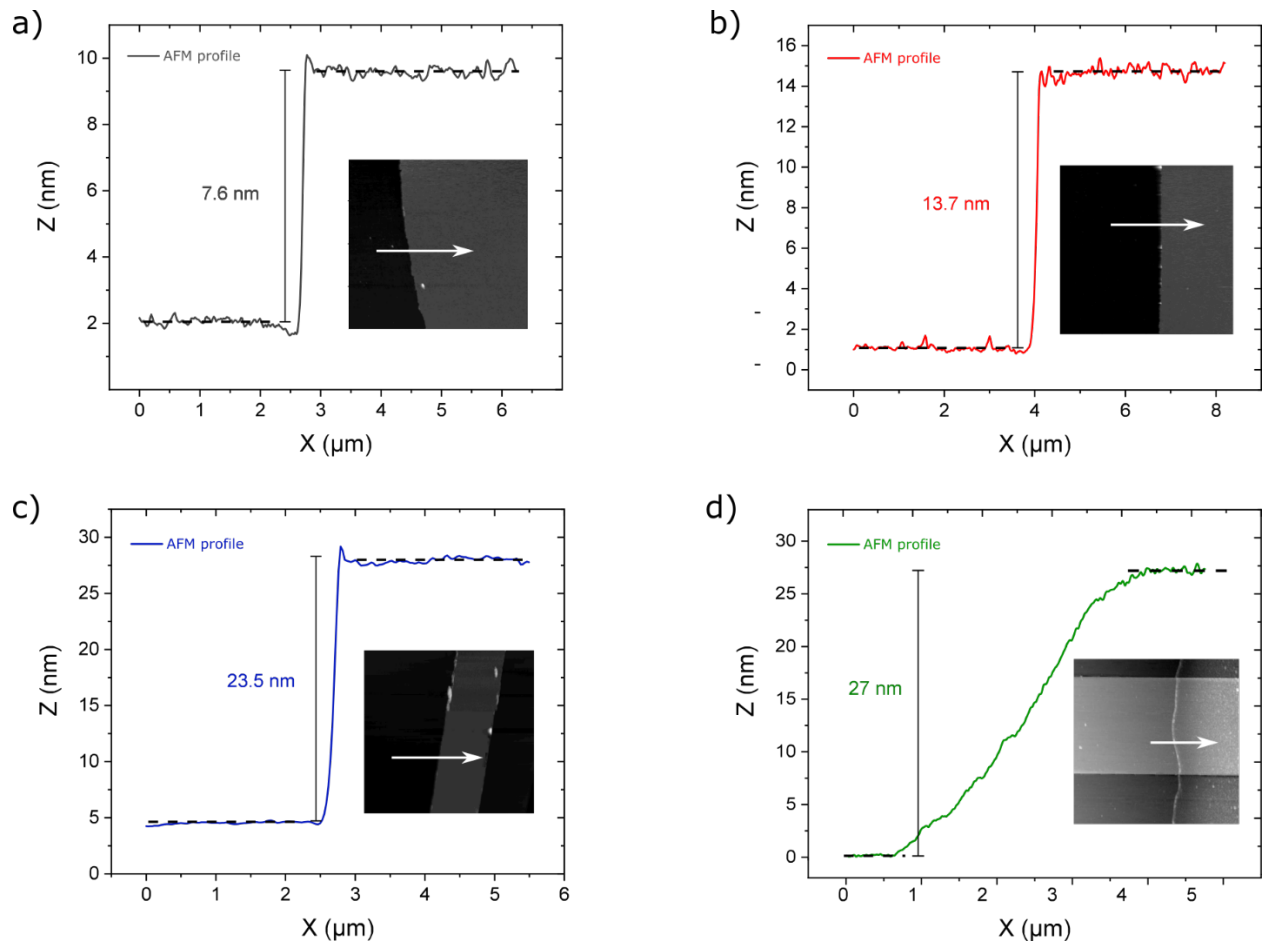

Figure S1. AFM Height measurements and corresponding AFM images of the PtSe$_2$ films.

**Opto-electrical measurement set up**

To measure the photoresponse of the photodetectors we have used a manual set-up shown schematically in Fig. S2. Laser light of a diode laser at 1550 nm wavelength were modulated at 1 kHz frequency using an electro-optic modulator (Thorlabs LN81S-FC) and guided above the grating couplers of the samples using a single-mode fiber. Photovoltages of the photodetectors were measured using a lock-in amplifier (Zurich Instrument MFLI). The bias voltages were applied to the detectors using a source meter (Keithley 2400). Photocurrents were calculated by dividing the photovoltages of the detectors by the total resistance of them.

All measured photodetectors have the same dimensions, 50 $\mu$m wide along the waveguide and 13 $\mu$m long perpendicular to the waveguide.

For time resolved measurements, the output of the photodetectors was connected directly to an oscilloscope (Zurich Instrument MFLI).

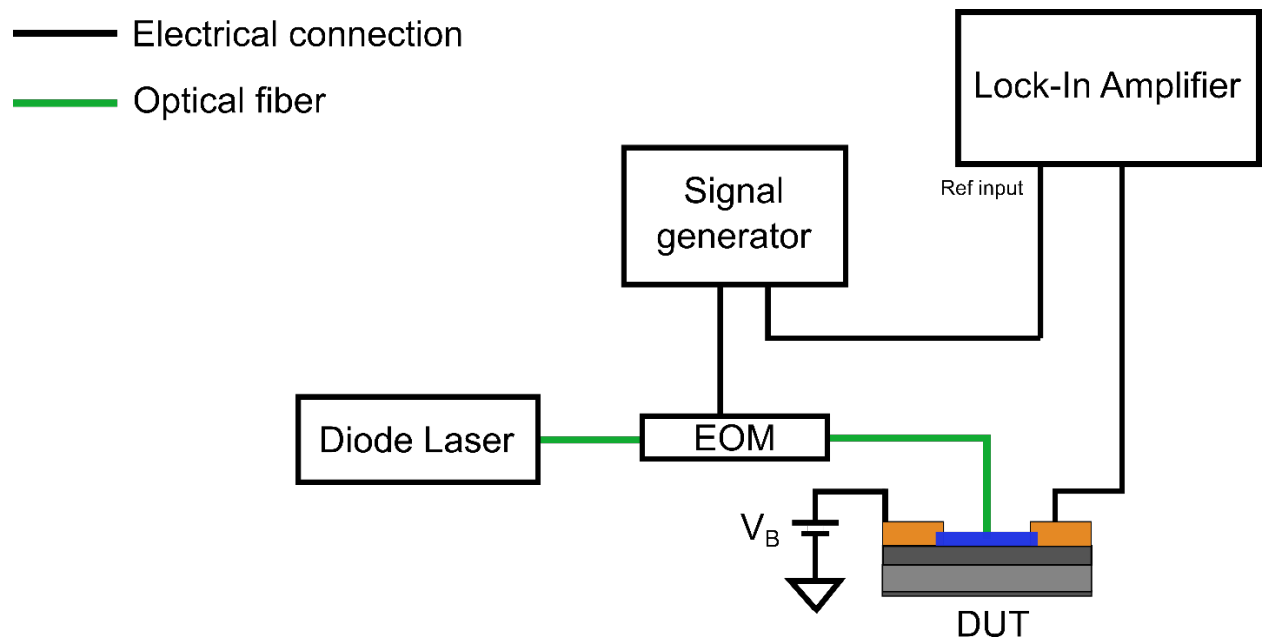

Figure S2. Schematic diagram of opto-electrical measurements set up.

## IV-Curve of PtSe₂ devices

$I_{ds}$-$V_{ds}$ measurements have been carried out on all multilayer PtSe$_2$ devices. The IV diagrams exhibit almost linear characteristics for drain voltages ranging from -100 mV to 100 mV (Fig. 2b). This indicates near-ohmic contacts between the films and the Ni/Al electrodes. When the drain voltages are increased, deviations from the linear behavior can be observed more clearly which indicate the existence of Schottky barriers at the interfaces. The reason for the presence of Schottky barriers at the interfaces may be due to a mix of semi-metallic and semiconducting crystallites and is subject to further studies.[1,2]

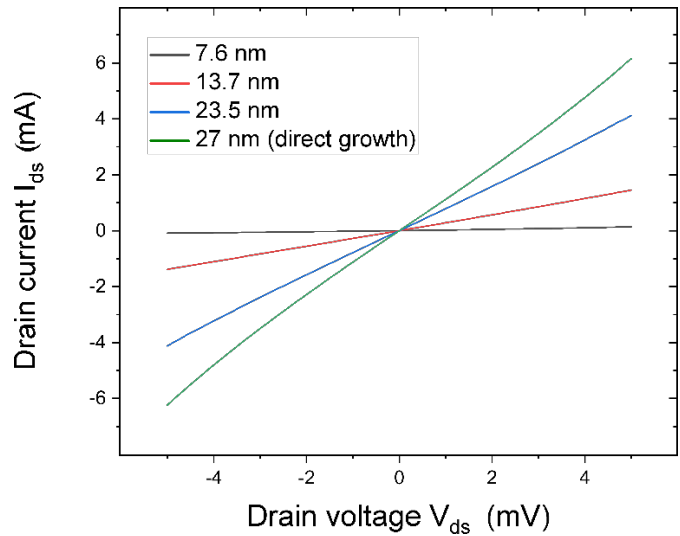

Figure S3. Drain current ($I_{ds}$) as a function of drain-source voltage ($V_{ds}$) for voltages ranging from -5 V to 5 V for all PtSe$_2$ devices.

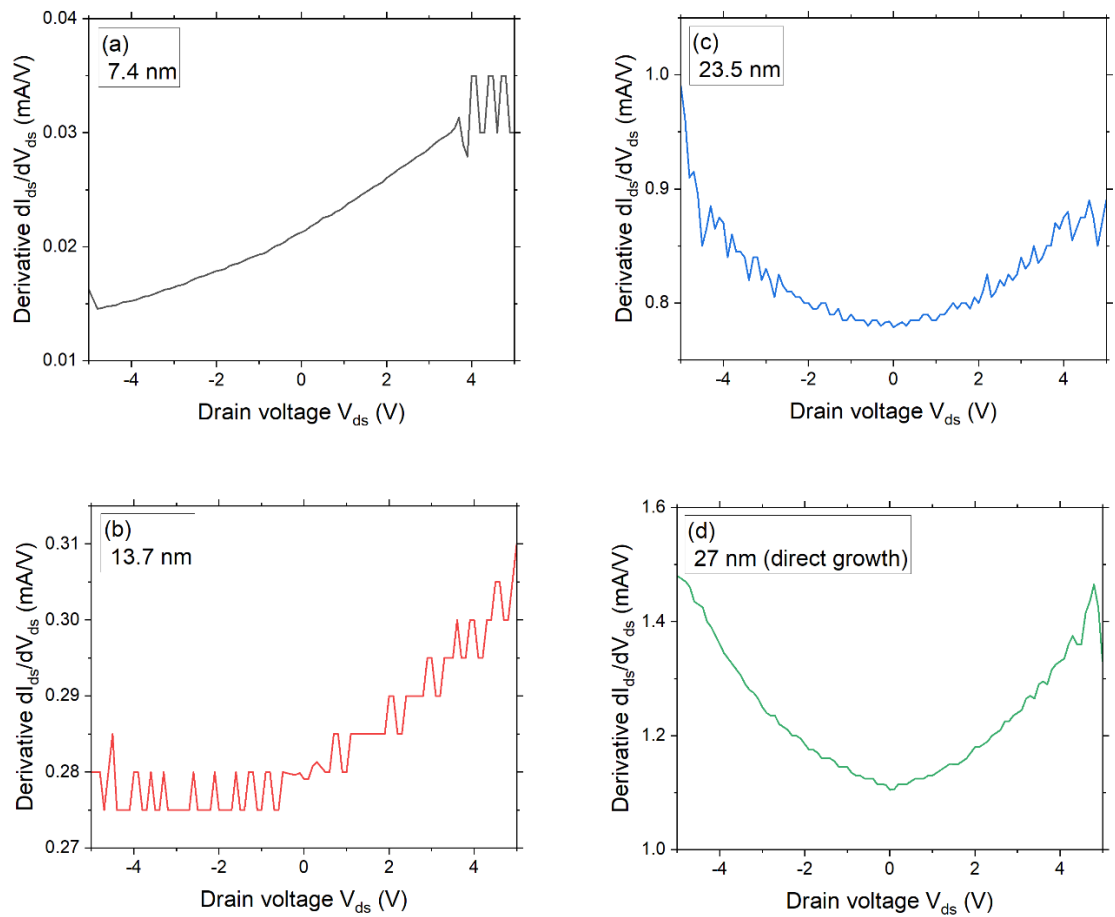

Figure S4. Derivatives of the drain current ($I_{ds}$) as a function of drain-source voltage ($V_{ds}$) for voltages ranging from -5 V to 5 V for all PtSe$_2$ devices revealing the non-linearity in the IV curves.